\documentclass[aps,reprint,eqsecnum, twocol,amsmath,amssymb]{revtex4-1}
\usepackage{graphicx}
\usepackage{float}
\usepackage{xcolor}

\newcommand{\p}{\partial}
\newcommand{\ep}{\varepsilon}
\newcommand{\om}{\omega}
\newcommand{\nn}{\nonumber}
\newcommand{\ta}{\theta}
\newcommand{\al}{\alpha}
\newcommand{\cN}{{\cal N}}
\newcommand{\cH}{{\cal H}}
\newcommand{\cF}{{\cal F}}
\newcommand{\cE}{{\cal E}}
\newcommand{\cW}{{\cal W}}
\newcommand{\sM}{M}
\newcommand{\wh}{\widehat}
\newcommand{\be}{\begin{equation}}                      
\newcommand{\ee}{\end{equation}}
\newcommand{\ba}{\begin{eqnarray}}
\newcommand{\ea}{\end{eqnarray}}
\newcommand{\bref}[1]{(\ref{#1})}
\newcommand{\js}{{js}}
\newcommand{\jo}{{jo}}
\newcommand{\je}{{je}}

\newcommand{\z}{\texttt 0}
\newcommand{\bi}[1]{\bibitem{#1}}

\begin{document}
\title{ Coupled-mode  theory 
	for microresonators with quadratic nonlinearity}

\author{Dmitry V. Skryabin}
\email{d.v.skryabin@bath.ac.uk}

\affiliation{Department of Physics, University of Bath, Bath BA2 7AY, UK}%

\begin{abstract}
We use Maxwell's equations to derive several models describing the interaction of the multi-mode fundamental field and its second harmonic in a ring microresonator with quadratic nonlinearity and quasi-phase-matching. We demonstrate how multi-mode three-wave mixing sums entering nonlinear polarisation response can be calculated via Fourier transforms of products of the field envelopes. Quasi-phase-matching gratings with arbitrary profiles are incorporated seamlessly into our models.    We also introduce several levels of approximations 
allowing to account for dispersion of nonlinear coefficients and demonstrate how coupled-mode equations can be reduced to the envelope Lugiato-Lefever-like equations with self-steepening terms. An estimate for the $\chi^{(2)}$ induced cascaded Kerr nonlinearity, in the regime of imperfect phase-matching, puts it above the intrinsic Kerr effect by several orders of magnitude.
\end{abstract}

	\maketitle
	

\section{Introduction}
Third-order, $\chi^{(3)}$, or Kerr, nonlinearity based frequency conversion, comb generation and soliton formation in high-quality factor  microresonators \cite{brag,vah,pas}  continue to be a fast-expanding direction of applied and fundamental research  in  small footprint nonlinear photonics \cite{rev1,rev2}. Using second-order, $\chi^{(2)}$,  nonlinearity  
has always existed as an alternative, and often a preferred  option, 
for  a variety of nonlinear optics applications \cite{rev3,rev4,rev5}. 
More than two decades ago Refs. \cite{old25,ronald,old26} reported experimental realisations, respectively, in space and time, of theoretical predictions \cite{old27,ronald0} of solitons due to quadratic nonlinearity, which has led to a surge of research in this area, and
that is when a sub-area of the resonator quadratic solitons has been shaped  \cite{rev3}. Though a viable for the resonator (cavity) solitons experimental setting has not been worked out at that time, basic three-wave mixing scenarios have been studied theoretically using adaptations of the  Lugiato-Lefever approach \cite{ll}. In particular, bright spatial and temporal solitons have been reported for the 
intra-cavity second-harmonic generation  \cite{old15},  degenerate \cite{old12,old13,ol} and non-degenerate \cite{old23} parametric down-conversion, including effects of different group velocities \cite{old24}.

 Refs. \cite{prl1,prl2} are the landmark papers on using high-quality factor whispering gallery microresonators with quadratic nonlinearity for frequency conversion applications. Foundational work on monolithic $\chi^{(2)}$ Fabry-Perot resonators has been accomplished back in 90th \cite{mon1,mon2}.   
Since then, this area has developed with a gradually accelerating pace,  see, e.g., Ref. \cite{rev4} for a few years old review
and Refs. \cite{micro9,micro4,micro6,micro5,micro1,micro8,japan,fins,micro3,soln6,micro2,loncar,nature,jan,alex2,alex,miro}
for some of the experimental contributions shaping its current state. There are a few good reasons for this surge of interest. Frequency combs with second harmonic are naturally octave wide and therefore are suitable for direct self-referencing \cite{micro7}. Quadratic nonlinearity is also relatively strong. Even if conversion into the second harmonic is inefficient, i.e., phase matching is not perfect, one still can derive advantages in terms of threshold powers
relative to the Kerr combs \cite{nature,jan,miro,alex}.  Modulational instability, i.e.,
side-band generation, due to $\chi^{(2)}$ effects happens through a range of both normal and anomalous dispersions \cite{rev3}, which comes handy, in particular, for visible and UV ranges, and should help to alleviate some of the design, material and pump laser constrains.
Material wise, lithium-niobate (LN) has been extensively used for both integrated and  
bulk cut  microresonator devices, see, e.g., \cite{rev4,rev5,loncar} and references there in. LN allows to utilise both birefringence-  and quasi- phase matching arrangements. Few other important materials to mention are silicon nitride, SiN,  \cite{micro4,micro9,guide1,guide2}, aluminium nitride, AlN, \cite{micro8,alex},  and gallium phosphate, GaP, \cite{micro2}, GaAs and AlGaAs \cite{optica,gla,bri}.

Regarding recent theory and modelling approaches to $\chi^{(2)}$ resonators,  
a significant effort \cite{bow2,bow6,han1} 
has been focused on utilising the Ikeda map method \cite{ikeda,mandel}
and its reductions to the mean field models  to
describe frequency conversion in centimetre and larger 
multi-mirror resonators with LN crystals used as a nonlinear element \cite{bow1,bow3,bow4,bow5,soln3}.
This approach yields a Lugiato-Lefever (LL)  model, where an evolution 
coordinate is a round-trip number, and a dispersion operator 
is applied in the time domain. Time and frequency are conjugate variables, in this formulation, which means that taking a Fourier transform of the envelope function 
across the computational domain periodic in time recovers frequency spectrum at a given round-trip. 

An alternative to the Ikeda map approach is to use a basis of the resonator modes to find a solution to the boundary value problem for the nonlinear and time-dependent Maxwell equations.   
This method has flourished after it has been applied to describe frequency comb generation 
and solitons in Kerr microresonators \cite{chembo1,chembo2,herr}. 
Ref. \cite{skr} uses an example a bidirectionally pumped Kerr 
resonator to provide a refreshed outlook on this formulation and includes a literature overview.
Coupled mode equations, under quite general assumptions, 
are formally equivalent to an LL equation \cite{skr}. 
This formulation of LL is, however, different from the one derived from the Ikeda map. It uses physical time as an evolution variable and its dispersion operator involves an angular coordinate \cite{chembo2,skr}. Hence, for a given time moment, 
it deals with a spectrum of mode numbers (momenta) and requires periodicity 
to be applied in the angular coordinate. Thus, this formulation of LL is naturally connected to the initial  and boundary conditions of Maxwell equations. Also, it  
links directly to the familiar methodology from the quantum mechanics text-books dealing with periodic boundary value problems for Schr\"odinger equation: $i\p_t\psi=-\tfrac{1}{2}D_2\p_\ta^2\psi+U(\ta)\psi$, $\psi(t,\ta)=\psi(t,\ta+2\pi)$. 
Here, polar angle, $\ta$, and momentum are the conjugate variables. This approach has an advantage of the transparent interpretation of, e.g., side-band growth, when periodic in space modes grow exponentially in time. Note, that a concern discussed in Ref. \cite{han2} related to comparison of the LL vs Ikeda approaches 
has been alleviated in the follow-up papers \cite{skr1} and \cite{fabio}, 
that used, respectively, the coupled-mode and Ikeda 
map connected formulations of the LL model.

Recently, a derivation of the coupled-mode and LL equations
from the Maxwell equations 
for a ring microresonator with $\chi^{(2)}$ nonlinearity has been sketched in Ref. \cite{soln7}. Refs. \cite{ingo1,alex} has also announced 
coupled-mode formulations. A relying on it (in fact, equivalent to it, see Section 8 below) 
LL model has 
been used in \cite{jan,soln7,soln4} to study $\chi^{(2)}$ combs and solitons, and its comparison with experimental data  has been  encouraging \cite{jan,alex}. 
Below we present a detailed derivation procedure of the coupled-mode equations from the Maxwell equations with $\chi^{(2)}$ nonlinearity and subsequently transform them to the envelope LL equations.  

One of our aims here is to present a formalism  underpinning a variety of models that has been and can be used by the community working on the frequency conversion and combs in $\chi^{(2)}$ microresonators. Apart from providing  methodological background for already advertised models  \cite{jan,soln7,soln4,ingo1,alex}, we also go beyond them in several aspects. First, we show how the mode number dependent variations of the nonlinear interaction strength for multimode three-wave mixing can be accounted for with various degrees of accuracy. Second, we formulate  coupled-mode equations by treating multimode nonlinear sums pseudo-spectrally. Here we use connections with 
pseudo-spectral approaches developed for quantum Schr\"odinger equation \cite{pseudo} and for Kerr microresonators \cite{skr,optcom,guo}. The pseudo-spectral method  provides a route for a seamless and approximation free incorporation of arbitrary quasi-phase-matching grating profiles and allows a user to apply fast-Fourier transform to deal with the multi-mode three-wave mixing nonlinear terms. Finally, we briefly address a problem of the relative strength of $\chi^{(2)}$ and $\chi^{(3)}$ effects.

Chapter titles used below are self-explanatory, 
which makes outlining their content here unnecessary.

\section{Maxwell equations and mode expansion}
Quite generally, Maxwell equations for an electric field component $ \cE_\al$ 
in a dispersive, spatially inhomogeneous and nonlinear material read as
\begin{subequations}
	\label{e2:1}
\begin{align}
&
 c^2\p_\al\p_{\al_1} \cE_{\al_1}-c^2\p_{\al_1}\p_{\al_1} \cE_\al \nn
\\ &+\p_t^2\int_{-\infty}^{\infty}\wh\ep_{\al\al_1}(t-t', r,\ta,z) \cE_{\al_1}(t',\vec r)dt'= -\p_t^2\cN_\al,
\label{e2:1a}\\
& \al=x,y,z,~ \al_1=x,y,z.
\label{e2:1b}
\end{align}
\end{subequations}
$x,y,z$ are  spatial coordinates, and implicit summations over the repeated $\alpha_1$ is assumed. The first and second terms in the left hand side of Eq. \bref{e2:1a} are gradient of divergence and Laplacian, respectively.
$\wh\ep_{\al\al_1}$ is the linear dielectric response varying in space and time, $t$.   	 $c$ is the vacuum speed of light.

To introduce a ring microresonator geometry, we proceed by defining $\ta=[0,2\pi)$ as the 
azimuthal coordinate varying along the ring circumference. $z$ axis is 
perpendicular to the ring plane, while $r=\sqrt{\smash[b]{x^2+y^2}}$ measures distance from the ring 
centre.  We assume that a microresonator is made of a $z$-cut uniaxial crystal
and $\wh\ep_{\al\al_1}$ is a diagonal matrix. Time to frequency Fourier transforms of  
$\wh\ep_{xx,yy}$ give  the ordinary refractive index, $n_o$, squared, and the transform of  
$\wh\ep_{zz}$ gives  the extraordinary index, $n_e$, squared. 

$\cN_\alpha$ is the second order nonlinear polarization,
\be 
\cN_\alpha=\chi^{(2)}_{\al\al_1\al_2}\cE_{\al_1} \cE_{\al_2}=2d_{\al\al_1\al_2}
\cE_{\al_1} \cE_{\al_2}, 
\label{e2:2}
\ee
where 
$d_{\alpha\alpha_1\alpha_2}$ is a reduced ($3\times 6$) tensor of the second 
order nonlinear susceptibility \cite{boyd,boyd1}. 
Linear, $\cN_{\al}=0$, modes of the resonator are divided 
into quasi- ordinarily polarised ones, $s=o$ (quasi-TE), and quasi- extraordinarily ones, 
$s=e$ (quasi-TM).   
Generally, modes of an open microresonator  have complex eigenfrequencies 
and do not form an orthogonal basis. We proceed to disregard these effects and later include loss and pump phenomenologically. 
 
$\cE_\al$ will be expressed few lines below as a superposition of the 
linear microresonator modes 
\begin{subequations}
	\label{e2:3}
\begin{align}
& \Theta_{s\al}(\ta)\Phi_{js\al}(r,z)e^{ij\ta- i\om_\js t},
\label{e2:3a}\\
&
 j\in \mathbb{Z}, j>0,
 	\label{e2:3b}\\
 & 
 s=o,e. 	\label{e2:3c} 
\end{align} 
\end{subequations}
Here
$j$ is an azimuthal mode number or angular momentum, and $\om_\js$ is 
the corresponding frequency. Linear modes can be calculated either asymptotically as Bessel functions \cite{modes1} or numerically using, e.g, COMSOL. In either case, they are commonly represented by a vector in the polar basis, while $\chi^{(2)}$ tensors are readily available in the literature in the Cartesian one.
Factorization of the intensity profiles into the $\ta$ 
and $(r,z)$ dependent parts and an assumption that the ordinary modes are quasi-transverse, i.e.,
their component tangential to the ring  can be neglected relative to the dominant radial one, are the approximations that we make here for the sake of transparency:
\begin{subequations}
	\label{e2:4}
	\begin{align}
\nn &\Theta_{ox}(\ta)=\cos\ta,~\Theta_{oy}(\ta)=\sin\ta,~\Theta_{oz}(\ta)=0,\\
&
\Phi_{jox}=\Phi_{joy}\equiv  \Phi_{jo}(r,z),
\label{e2:4a}
\\  
\nn &\Theta_{ex}(\ta)=\Theta_{ey}(\ta)=0,~\Theta_{ez}(\ta)=1,\\
&
\Phi_{jez}\equiv  \Phi_{je}(r,z).
\label{e2:4b}
	\end{align}
	\end{subequations}
Here Eqs. \bref{e2:4a} and \bref{e2:4b} apply, respectively, to the families 
of the ordinary and extraordinary modes. 

Three electric field  components $\cE_\al$ 
can be expressed via the ordinary and extraordinary fields, $\cE_{o,e}$, as
\begin{subequations}
	\label{e2:5}
	\begin{align}
 & \cE_x=\cos\ta~\cE_o,\label{e2:5a} \\
 & \cE_y=\sin\ta~\cE_o,\label{e2:5b} \\
 & \cE_z=\cE_e,
 \label{e2:5c}
 \\
& \cE_s=\sum_{j=j_{\min,s}}^{j_{\max,s}}b_{js} 
\Phi_{js}B_{js}(t)e^{ij\ta-i\om_{js}t}+
c.c.,
\label{e2:5d}
\end{align}
\end{subequations}
where $B_\js$ are the complex mode amplitudes and $b_\js$ are the normalisation constants.  We choose to normalise linear modes as 
\begin{equation}
\max_{r,z} |\Phi_{js}|=1,
\end{equation}
 and hence units of  $b_\js B_\js$ are V/m.  
Scaling factors $b_\js$ are defined to  measure
$|B_{\js}|^2$  in Watts: 
\be 
b_\js^2=\frac{{\cal Z}_{vac}}{2S_\js n_{sj}},
\label{e2:6}
\ee
where ${\cal Z}_{vac}=1/\epsilon_{vac}c=377$ V$^2$/W is the free space impedance \cite{boyd}.   
$S_\js=\big(\iint |\Phi_{js}^\prime|^2 dxdz\big)^{2}$  $\big(\iint |\Phi_{js}^\prime|^4 dxdz\big)^{-1}$ 
is the 
effective transverse mode area,
$\Phi_{js}^\prime=\Phi_{js}(r\vert_{y=0},z)$.
Refractive indices  are  
$n_{oj,ej}^2=\int\wh\ep_{xx,zz}(\tau,r=r_0,z=0)e^{i\om_{jo,je}\tau}d\tau$,
where $r_0$ is the distance from the resonator axis to the intensity maximum.  

Below, we consider in details an example when 
the ordinary modes are grouped around $j=M_o$ corresponding to the frequency 
$\om_{\sM_o}$ and the extraordinary ones around $j=M_e$ with $\om_{\sM_e}$. 
 If the pump is ordinarily polarised, then $\om_{\sM_e}\simeq 2\om_{\sM_o}$ is its  second harmonic. 
Resonance frequencies of the ordinary and extraordinary modes can be approximated as
\begin{subequations}
	\label{e2:7}
	\begin{align}
\nn &\om_{js}=\om_{\sM_s}+ D_{1s}(j_s-M_s)+\tfrac{1}{2!}D_{2s}(j_s-M_s)^2\\ &
+\tfrac{1}{3!}D_{3s}(j_s-M_s)^3+\tfrac{1}{4!}D_{4s}(j_s-M_s)^4+\dots,
\label{e2:7a}\\
&\nn \\
 &j_s=\dots M_s-2,M_s-1,M_s,M_s+1,M_s+2,\dots
\label{e2:7b}
\end{align}
\end{subequations}
$D_{1s}/2\pi$ are the repetition rate parameters equalling the free spectral ranges (FSRs).
$D_{2s}$ are the second-order dispersions and $D_{(k>2)s}$ are the higher-order dispersion coefficients. We consider the same number, $N$, of modes around both $M_o$ and $M_e$, and therefore can introduce the 
same mode number offset parameter $\mu$,
\be \label{e2:8}
	\mu=-\tfrac{1}{2}N+1,\dots,\z,\dots,\tfrac{1}{2}N,	
\ee
giving
\begin{subequations}
	\label{e2:9}
	\begin{align}
\label{e2:9a}
& \om_{\mu s}=\om_{\z s}+ 
D_{1s}\mu+\tfrac{1}{2!}D_{2s}\mu^2+\tfrac{1}{3!}D_{3s}\mu^3+\dots,\\
& \om_{(j=\sM_s) s}=\om_{(\mu=\z) s},
\label{e2:9b}
\end{align}
\end{subequations}
for the ordinary and extraordinary resonances.

\section{Coupled-mode equations and chi-2 tensor}
Making a substitution  $t'=t-\tau$ in Eqs. \bref{e2:1a}, \bref{e2:5b} 
we then assume that material response is fast so that 
$B_{js}(t-\tau)\simeq B_{\js}(t)-\tau\p_t B_{\js}+\dots$. 
Neglecting all the 2nd and higher-order time derivatives of  $B_\js$ we find that Eq. \bref{e2:1a} transforms to 
\begin{align}
&
\nn c^2\p_\al\p_{\al_1} \cE_{\al_1}-c^2\p_{\al_1}\p_{\al_1} \cE_\al
+\p_t^2\int_{-\infty}^{\infty}\wh\ep_{\al\al_1}(t-t',\vec r) \cE_{\al_1}(t',\vec r)dt' \\ 
 & \simeq\sum_\js b_\js\Theta_{s\al}\Phi_\js
\Big(-2i\om_\js  n^2_{s j} 
e^{ij\ta-i\om_\js t}\p_tB_\js+c.c.\Big)=-\p_t^2\cN_\al.
\label{e3:1}
\end{align}
Refractive indices $n^2_{s j}$ inside the round bracket approximate
$(n^2_{s j} +\frac{1}{2}\om_{js}\p_\om n^2_{sj})$ combinations formally 
emerging there. 

 Eqs. \bref{e3:1} are then projected on the linear modes, Eqs. \bref{e2:3}, \bref{e2:4},
\be
 i \p_tB_\js=\frac{e^{i\om_\js t}}{2\om_\js b_\js^2n_\js^2V_\js}\p_t^2\int_0^{\infty} rdr 
 \int_{-\infty}^{\infty} dz  \int_{0}^{2\pi} d\ta
 \cN_s b_\js \Phi_\js e^{-ij\ta}.
\label{e3:2}
\ee
Here 
\begin{equation}
V_\js=2\pi\iint \Phi_\js^2~rdrdz 
\end{equation}
are the  mode volume parameters,
 and 
\begin{subequations}
	\label{e3:3}
	\begin{align}
& \cN_o=\cos\ta \cN_x+\sin\ta \cN_y,
\label{e3:3a}\\
& \cN_e= \cN_{z}
\label{e3:3b}
\end{align}
\end{subequations}
are the ordinary and extraordinary  nonlinear polarizations.

We choose to consider   LiNbO$_3$ crystal as an example. It
belongs to the point group $3m(C_{3v})$,  and its 
nonlinear polarization vector  \cite{boyd,boyd1} is $\left(
	\cN_x, \cN_y, \cN_z\right)^T=G(\ta)\times$ 
\be
2\left[\begin{array}{cccccc}
0&0&0&0&d_{31}&-d_{22}\\
-d_{22}&d_{22}&0&d_{31}&0&0\\
d_{31}&d_{31}&d_{33}&0&0&0
\end{array}\right]\left(\begin{array}{c}
	\cE_x^2\\
	\cE_y^2\\
	\cE_z^2\\
	2\cE_y\cE_z\\
	2\cE_x\cE_z\\
	2\cE_x\cE_y
\end{array}\right).
\ee
Hence,
\begin{subequations}
	\label{e3:4}
	\begin{align}
 &
 \cN_x=2G(\ta)\big(2d_{31}\cE_z\cE_x-2d_{22}\cE_x\cE_y\big),
 \label{e3:4a}
  \\ & 
 \label{e3:4b} 
 \cN_y=2G(\ta)\big(-d_{22}\cE_x\cE_x+d_{22}\cE_y\cE_y+2d_{31}\cE_y\cE_z\big),
 \\
 &\label{e3:4c}
 \cN_z=2G(\ta)\big(d_{31}\cE_x\cE_x+d_{31}\cE_y\cE_y+d_{33}\cE_z\cE_z\big).
\end{align}
\end{subequations}
Values of the nonlinear tensor elements are
$d_{22}\simeq 2.3$pm/V, $d_{31}\simeq 4.8$pm/V, $d_{33}\simeq 29.7$pm/V \cite{boyd1}.
$G(\ta)$ is a profile of the quasi-phase matching grating \cite{rev4,qpm}, see Section 4. 

Using Eqs. \bref{e2:5a}, \bref{e3:4} helps to transform Eqs. \bref{e3:3}
\begin{subequations}
	\label{e3:5}
	\begin{align}
	&\cN_o=2G(\ta)\big(2d_{31}~\cE_o\cE_e-d_{22}
\sin 3\ta~\cE_o^2\big),
\label{e3:5a}
 \\ & 
 \label{e3:5b} 
 \cN_e=2G(\ta)\big(d_{31}\cE_o^2+d_{33}\cE_e^2\big).
\end{align}
\end{subequations}

Eqs.  \bref{e3:2} considered together with Eqs. \bref{e3:5}, \bref{e2:5d} make a self-consistent system of equations, that only miss  appropriate pump and loss terms to model a typical frequency conversion experiment in a microresonator. Postponing
introducing of the latter saves us a considerable amount of space in Section 5, where we elaborate approximations for Eqs. \bref{e3:2} allowing to efficiently 
handle integration in $r$ and $z$, while still accounting for  dispersion of nonlinear effects.

\section{Momentum and frequency matching}
The efficiency of frequency conversion from the pump to the desired frequency 
is dominantly driven by an interplay of the angular momentum matching and 
linear dispersion. The latter is encoded  in $\om_\js$ and becomes important for the multi-mode three-wave mixing, i.e., comb generation. The angular momentum matching is user adjusted
via, e.g, either temperature tuning of birefringence 
or periodic polling (quasi-phase-matching) \cite{rev4,qpm,modes5}.  

We assume that the quasi-matching grating is a square  function with  single or multiple periods and oscillating between $-1$ and $+1$,
\begin{subequations}
	\label{e4:1}
\begin{align}
& G(\ta)=\sum_{m=-\infty}^{\infty}G_me^{im\ta},
\label{e4:1a}
\\ 
& \max_m G_m=G_{\pm g}, g\in\mathbb{Z},
\label{e4:1b}
\end{align}
\end{subequations}
where $m$ are the momenta of the grating harmonics.
$|m|=g$ provides the leading order momentum matching conditions, with
$2\pi r_0/|g|$ being the corresponding grating period, see Eq. \bref{e4:2} below.
LiNbO$_3$ can be either quasi-phase-matched or birefringence phase-matched \cite{rev4}. 
The latter case has been used, e.g., to achieve recent microresonator comb results in the second harmonic configuration \cite{jan,miro}. 

We consider in details an example when the momentum matching is
arranged between the 
ordinary mode $j=M_o$ at a lower frequency $\om_{\z o}$
and the extraordinary mode $j=M_e$ with $\om_{\z e}$ that matches $2\om_{\z o}$ 
either exactly or as close as possible.
Taking momentum matching as
\be
M_e=2M_o+g,
\label{e4:2}
\ee
$g$ has to be chosen to minimize (ideally, to make exactly zero) 
absolute value of the  frequency mismatch, $\ep$, defined as
\be
\ep=  2\om_{\z o}-\om_{\z e},
\label{e4:3}
\ee
recall notational transformation from $\om_{js}$ to $\om_{\mu s}$ in Eqs. 
\bref{e2:9}.
Birefringence based momentum  matching, i.e., $G(\ta)=1$, $g=0$,  is, obviously, 
$M_e=2M_o$. 

The above phase matching considerations, as well as 
formalism developed here, can also 
be applied for $d_{33}$ mediated three-wave mixing 
processes involving only extraordinary resonances, 
and for  the ordinary resonances interacting via $d_{22}$, 
and for any combination of thereof, see Eqs. \bref{e3:5}.

\section{Tensors of  nonlinear coefficients and frequency mismatches}
If second harmonic generation from the ordinary to extraordinary modes 
is frequency and momentum matched, 
then $d_{31}$ element is engaged for both $\cN_o$ and $\cN_e$, and the non-matched,  $d_{22}$ (ordinary to ordinary conversion) and $d_{33}$ (extraordinary to extraordinary conversion)  mediated three-wave mixing processes can be disregarded. So that, Eqs. \bref{e3:2} become
\begin{subequations}
\label{e5:1}
\begin{align}
& i \p_tB_\jo= \frac{2e^{i\om_\jo t}}{\om_\jo b_\jo n_\jo^2V_\jo}\p_t^2
\iint rdrdz~ d_{31}\Phi_\jo\int d\ta~ G e^{-ij\ta}\cE_o \cE_e,
\label{e5:1a}
\\ 
& i \p_tB_\je=\frac{e^{i\om_\je t}}{\om_\je b_\je n_\je^2V_\je}\p_t^2
\iint rdrdz~d_{31}\Phi_\je \int d\ta~ G e^{-ij\ta} \cE_o\cE_o.
\label{e5:1b} 
\end{align}
\end{subequations}
We then express $\cE_s$ via their respective mode expansions, Eqs. \bref{e2:5b}, and
replace $j$ with $\mu$, see Eqs. \bref{e2:7}, \bref{e2:8}, \bref{e2:9}, which yields double sum expressions for $\cE_o\cE_e$ and $\cE_o^2$, 
\begin{subequations}
	\label{e5:2}
	\begin{align}
i \p_t B_{\mu o}=& \frac{2e^{i\om_{\mu o} t}}{\om_{\mu o} b_{\mu o} n_{\mu o}^2V_{\mu o}}\p_t^2
\iint rdrdz~ d_{31}\Phi_{\mu o}\int d\ta~ G  e^{-i(\mu+M_o)\ta} 
\nn  
\\
\times\sum\nolimits_{\mu_1\mu_2}
&\big(b_{\mu_1 o}\Phi_{\mu_1 o}B_{\mu_1 o}e^{i(\mu_1+M_o)\ta-i\om_{\mu_1 o}t}+c.c.\big)
\nn 
\\
\times &
\big(b_{\mu_2 e}\Phi_{\mu_2 e}B_{\mu_2 e}e^{i(\mu_2+M_e)\ta-i\om_{\mu_2 e}t}+c.c.\big),  
\label{e5:2a}
\\
i \p_tB_{\mu e}=& \frac{e^{i\om_{\mu e} t}}{\om_{\mu e} b_{\mu e} n_{\mu e}^2V_{\mu e}}\p_t^2
\iint rdrdz~ d_{31}\Phi_{\mu e}\int d\ta~ G  e^{-i(\mu+M_e)\ta} 
\nn 
\\
\times\sum\nolimits_{\mu_1\mu_2}&
\big(b_{\mu_1 o}\Phi_{\mu_1 o}B_{\mu_1 o}e^{i(\mu_1+M_o)\ta-i\om_{\mu_1 o}t}+c.c.\big)
\nn
\\
\times 
&\big(b_{\mu_2 o}\Phi_{\mu_2 o}B_{\mu_2 o}e^{i(\mu_2+M_o)\ta-i\om_{\mu_2 o}t}+c.c.\big).  
\label{e5:2b}
\end{align}
\end{subequations}
	We recall here, that, $\Phi_{\mu o}$ and $\Phi_{\mu e}$ with the same $\mu$
	refer to different $j$, $j=M_o+\mu$ and $j=M_e+\mu$, respectively, see Eqs. 
	\bref{e2:7}, \bref{e2:8}. Same applies to all other variables and 
	parameters.
	
By construction, the extraordinary frequencies  cluster around the second harmonics of the  ordinary ones  and therefore, while opening up the brackets,  we account only for the terms oscillating with frequencies $\om_{\mu_2 e}-\om_{\mu_1 o}$ in Eq. \bref{e5:2a} and for 
$\om_{\mu_2 o}+\om_{\mu_1 o}$ in Eq. \bref{e5:2b},  and disregard the rest,
\begin{subequations}
	\label{e5:3}
\begin{align} 
i \p_tB_{\mu o}\simeq & -\frac{2e^{i\om_{\mu o} t}}{\om_{\mu o} b_{\mu o} n_{\mu o}^2V_{\mu o}}
\nn \\&\times \iint rdrdz~ d_{31}\Phi_{\mu o}\int d\ta~ G  e^{i(M_e-2M_o)\ta}
	\nn \\
& 
\times\sum\nolimits_{\mu_1\mu_2}(\om_{\mu_2 e}-\om_{\mu_1 o})^2
b_{\mu_1 o}b_{\mu_2 e}\Phi_{\mu_1 o}\Phi_{\mu_2 e}
\nn
\\
&\times B^*_{\mu_1 o}B_{\mu_2 e}e^{i(\mu_2-\mu_1-\mu)\ta-i(\om_{\mu_2 e}-\om_{\mu_1 o})t},
\label{e5:3a}\\
i \p_tB_{\mu e}\simeq & -\frac{e^{i\om_{\mu e} t}}{\om_{\mu e} b_{\mu e} n_{\mu e}^2V_{\mu e}}\nn\\
&\times
\iint rdrdz~ d_{31}\Phi_{\mu e}\int d\ta~ G  e^{i(2M_o-M_e)\ta}
\nn \\
& 
\times\sum\nolimits_{\mu_1\mu_2}(\om_{\mu_2 o}+\om_{\mu_1 o})^2
b_{\mu_1 o}b_{\mu_2 o}\Phi_{\mu_1 o}\Phi_{\mu_2 o}\nn
\\
&\times B_{\mu_1 o}B_{\mu_2 o}e^{i(\mu_2+\mu_1-\mu)\ta-i(\om_{\mu_2 o}+\om_{\mu_1 o})t}.
\label{e5:3b}
\end{align}
\end{subequations}
Using Eq. \bref{e4:1} and after integrating in $\ta$,  
the only non-zero terms left in the right-hand sides of Eqs. \bref{e5:3} 
are the ones satisfying an extended set of momentum matching conditions:
\begin{subequations}
	\label{e5:4}
	\begin{align}
 \text{difference frequency generation, $(eo\to o)$:}&\nn \\
 \mu_2-(\mu_1-g)+m=\mu,&
\label{e5:4a}\\
 \text{sum frequency generation, $(oo\to e)$:}&\nn \\
 \mu_2+(\mu_1-g)+m=\mu,&
\label{e5:4b}
\end{align}
\end{subequations}
and describing all accounted for three wave mixing processes.
Each of these frequency sum and frequency difference processes has its
own frequency mismatch parameter,
\begin{subequations}
	\label{e5:5}
	\begin{align}
	\text{difference frequency generation:}&\nn \\
	\ep^{eoo}_{\mu_2\mu_1\mu, m}=\om_{\mu_2 e}-\om_{\mu_1 o}-\om_{\mu o},&
	\label{e5:5a}\\
	\text{sum frequency generation:}&\nn \\
	\ep^{ooe}_{\mu_2\mu_1\mu, m}=\om_{\mu_2 o}+\om_{\mu_1 o}-\om_{\mu e}.&
\label{e5:5b}
	\end{align}
\end{subequations} 
Here and below an index in the top row of indices indicates polarization of a 
photon with  the  respective momentum index  positioned in the bottom row. 
Thus, $\ep^{s_2s_1s}_{\mu_2\mu_1\mu, m}$
are the the rank-4 frequency mismatch tensors in $\mu$ and $m$.
Frequency and momentum matching considerations in Section 4, see Eqs. 
\bref{e4:2}, \bref{e4:3}, map onto the frequency mismatch tensors  as  
$\ep^{ooe}_{\z\z\z, g}=-\ep^{eoo}_{\z\z\z,-g}=\ep$. 
 
After integrals are taken, Eqs. \bref{e5:3} become a system of the ordinary 
differential equations for $B_{\mu s}$,
\begin{subequations}
	\label{e5:6}
	\begin{align}
	 i \p_t B_{\mu o}=&
	-	\sum_{\mu_1\mu_2 m}\gamma_{\mu_2\mu_1\mu, 
	m}^{eoo}\wh\delta_{\mu_2,\mu+\mu_1-g-m}\nn \\ &
		\times G_m B^*_{\mu_1 o} B_{\mu_2 e} 
		\exp\{-it\ep^{eoo}_{\mu_2\mu_1\mu, m}\},
	\label{e5:6a}\\
	i \p_tB_{\mu e}=&-
	\sum_{\mu_1\mu_2 m}\gamma_{\mu_2\mu_1\mu, 
	m}^{ooe}\wh\delta_{\mu_2,\mu-\mu_1+g-m}\nn \\ &
	\times G_m B_{\mu_1 o}B_{\mu_2 o}
	\exp\{-it\ep^{ooe}_{\mu_2\mu_1\mu, m}\}.
	\label{e5:6b}
	\end{align}
\end{subequations}
Here the Kronecker symbol $\wh\delta_{\mu_2,\mu_1}=1$ for $\mu_1=\mu_2$, and is 
zero otherwise. Combinations of indices in $\wh\delta$'s inside Eqs. 
\bref{e5:6} express the momentum matching conditions in Eqs. \bref{e5:4}. 
Nonlinear coefficients $\gamma_{\mu_2\mu_1\mu, m}^{s_2s_1s}$ are the rank-4 
tensors with the same index arrangements as in the frequency mismatch tensors,
\begin{subequations}
	\label{e5:7}
	\begin{align}
	 \gamma_{\mu_2\mu_1\mu, m}^{eoo}=& 
	\frac{2\big(\om_{\mu_{2} e}-\om_{\mu_{1} o}\big)^2}{\om_{\mu o}}~\frac
	{ b_{\mu_{2} e}b_{\mu_{1} o}}
	{b_{\mu o} n_{\mu o}^2 }\nn \\&\times
	 \frac{2\pi}{V_{\mu o}}\iint rdrdz~ d_{31} \Phi_{\mu_{2} e} 
	\Phi_{\mu_{1} o} \Phi_{\mu o}, 
	\label{e5:7a}\\
	 \gamma_{\mu_2\mu_1\mu, m}^{ooe}=&\frac{\big(\om_{\mu_{2} o}+\om_{\mu_{1} 
	 	o}\big)^2}{\om_{\mu e}}~\frac{b_{\mu_{2} o} 
		b_{\mu_{1} o}}{b_{\mu e} n_{\mu e}^2}\nn\\ &\times
	 \frac{2\pi}{V_{\mu e}}\iint rdrdz~d_{31}\Phi_{\mu_{2} o}\Phi_{\mu_{1} o} 
	 \Phi_{\mu e}.   
	\label{e5:7b}
	\end{align}
\end{subequations}

To fully set parameters for Eqs. \bref{e5:6}, one has to specify elements of 
$\ep^{s_2s_1s}_{\mu_2\mu_1\mu, m}$ and $\gamma_{\mu_2\mu_1\mu, m}^{s_2s_1s}$. 
Momentum matching conditions \bref{e5:4} fix, e.g., $\mu_2$, thereby reducing 
the number of nonzero tensor elements. For every quasi-phase matching order, 
$m$, one still  
needs to fix  $\sim N^2$ constants.
Doing so is not a problem for the frequency offset tensors, but knowing every 
$\gamma_{\mu_2\mu_1\mu m}^{s_2s_1s}$ takes calculating a two dimensional 
integral. This task becomes impractical already for $N\sim 10^2$.
However, if the research with $\chi^{(2)}$ microresonators is to progress to the generation of the octave and wider combs, and to designing of emission of resonance radiation, see, e.g., \cite{micro1,oct1,oct2,oct3}, then dispersion of nonlinear interaction is desirable to account for. 

One approach to account for changes in values of $\gamma_{\mu_2\mu_1\mu, 
m}^{s_2s_1s}$ follows from the consideration, that initiation of spectral 
broadening happens through those $B_{\mu_1 s_1}B_{\mu_2 s_2}$ products, where 
one, or both, of the participating photons is the pump one.  Thus, fixing 
$\mu_1=0$ and using momentum matching to find $\mu_2$ give $\mu_2=\mu-m-g$ for 
the frequency difference terms and $\mu_2=\mu-m+g$ for the sum-frequency ones. 
$m=\pm g$ is also a natural choice. Hence,  using a substitution  
\begin{subequations}
	\label{e5:8}
	\begin{align}
	& \gamma_{\mu_2 \mu_1 \mu, m}^{eoo}\to\gamma_{\mu \z \mu, -g}^{eoo}\equiv 
	\gamma_{\mu o},
	\label{e5:8a}	\\ 
&	\gamma_{\mu_2 \mu_1 \mu, m}^{ooe}\to\gamma_{\mu \z \mu, g}^{ooe}\equiv 
\gamma_{\mu e}   
	\label{e5:8b}
	\end{align}
\end{subequations}
allows to retain main features of the  
dispersion of $\gamma_{\mu_2\mu_1\mu, m}^{s_2s_1s}$  coefficients.

Eqs. \bref{e5:6} now become  
\begin{subequations}
	\label{e5:9}
	\begin{align}
	 i \p_t B_{\mu o}&=
	-\gamma_{\mu o}
	\sum_{\mu_1\mu_2 m}\wh\delta_{\mu_2,\mu+\mu_1-g-m}
	G_m B^*_{\mu_1 o} B_{\mu_2 e} \nn\\
&\times	\exp\{-it\ep^{eoo}_{\mu_2\mu_1\mu, m}\},
	\label{e5:9a}\\
		i \p_tB_{\mu e}&=-\gamma_{\mu e}
	\sum_{\mu_1\mu_2 m}\wh\delta_{\mu_2,\mu-\mu_1+g-m}
	G_m B_{\mu_1 o}B_{\mu_2 o}\nn\\
&\times	\exp\{-it\ep^{ooe}_{\mu_2\mu_1\mu, m}\}.
	\label{e5:9b}
	\end{align}
\end{subequations}
Here, the number of nonlinear coefficients to be set is $2N$ and 
the number of  the respective double
integrals to be taken is $N$.

Let us introduce more approximations for the sake of 
elaborating  more explicit forms for  $\gamma_{\mu s}$.
We first  use Eq. \bref{e2:9}  and approximate 
$\om_{\mu e}-\om_{\z o}= \omega_{\z o}+D_{1e}\mu+{\cal O}(D_{2s}\mu^2)$, 
$\om_{\mu o}+\om_{\z o}= 2\om_{\z o}+D_{1o}\mu+{\cal O}(D_{2s}\mu^2)$.
Thus,
\begin{subequations}
	\label{e5:10}
	\begin{align}
	\gamma_{\mu o}\simeq& 
	2\om_{\z o}\left(1+\frac{2D_{1e}\mu}{\omega_{\z o}}\right)
	\nn \\&\times
	\frac	{ b_{\mu e}b_{\z o}}
	{b_{\mu o} n_{\mu o}^2 }
	\frac{2\pi}{V_{\mu o}}\iint rdrdz d_{31}\Phi_{\mu e} 
	\Phi_{\z o} \Phi_{\mu o}, 
	\label{e5:10a}\\
	\gamma_{\mu e}\simeq&2\om_{\z o}\left(1+\frac{D_{1o}\mu}{\omega_{\z o}}\right)\nn \\&\times
	\frac{b_{\mu o} b_{\z o}}{b_{\mu e} n_{\mu e}^2}
	\frac{2\pi}{V_{\mu e}}\iint rdrdz d_{31}\Phi_{\mu o}\Phi_{\z o} 
	\Phi_{\mu e}.   
	\label{e5:10b}
	\end{align}
\end{subequations}
Eqs. \bref{e5:10} account for dispersion of nonlinear coefficients that originates in the frequency pre-factors, see round brackets, 
and for dispersion of the mode profiles within the modal group around the pump and second harmonic fields, see expressions after '$\times$'.
Pump (ordinary) and second harmonic (extraordinary) are separated by an octave, therefore the mode profile change between them can be significant. However, the mode profile changes within the  ordinary and extraordinary spectra can be disregarded to facilitate further transparency of the coefficient structure,
	\begin{align}
	\gamma_{\mu o}\simeq\gamma_{\z o} 
	\left(1+\frac{2D_{1e}\mu}{\omega_{\z o}}\right), ~
	\gamma_{\mu e}\simeq\gamma_{\z e}\left(1+\frac{D_{1o}\mu}{\omega_{\z o}}\right).   
	\label{e5:11}
	\end{align}
Eqs. \bref{e5:11} represent a reasonable approximation, that captures the linear in $\mu$ change of nonlinear interaction strength. In the real space this dependence is associated with the self-steepening effects, see Section 8A. It becomes more important for shorter resonators with higher repetition rates. If $D_{1s}/2\pi\sim 0.1$THz then $\mu\sim 100$ makes $D_{1s}\mu/\om_{\z o}\sim 0.1$, which  complies with approximations of Eqs. \bref{e5:11} and, at the same time, makes dispersion of nonlinearity appreciable. Taking  $D_{1s}/2\pi\sim 1$THz  suggests using  Eqs. \bref{e5:8}, \bref{e5:7}. 

To get $\gamma_{\z s}$ in Eqs. \bref{e5:11} one should use Eqs. \bref{e5:10}.  
Integrals there can be roughly approximated using 
 Gaussian mode profiles and  $rdr\approx r_0dx$: $2\pi\iint \Phi_{\z o}^2\Phi_{\z e} rdrdz/V_{\z s}\simeq\tfrac{2}{3}$. Hence,  $\gamma_{\z s}$ are estimated as
\be
\gamma_{\z o}\approx \frac{4d_{31}\om_{\z o}b_{\z e}}{3n_{\z o}^2},~
\gamma_{\z e}\approx \frac{4d_{31}\om_{\z o}b_{\z o}^2}{3n_{\z e}^2b_{\z e}}.
\label{e5:12}
\ee
Recalling Eq. \bref{e2:6} and taking $n_{\z s}=2.2$, $S_{\z o}=50\mu$m$^2$, $S_{\z e}=30\mu$m$^2$
we find $b_{\z o}\simeq1.3\cdot10^6$, $b_{\z e}\simeq1.7\cdot10^6~$W$^{-1/2}$V/m.
Then, $d_{31}=4.8$pm/V and  
$\om_{\z o}/2\pi= 200$THz give  $\gamma_{\z o}/2\pi\approx 350$,
$\gamma_{\z e}/2\pi\approx 270~$MHz/W$^{1/2}$.

\section{Pump and loss}
We assume that a microresonator is pumped into a single ordinarily polarised mode, 
$\mu=0$, and that the laser frequency  $\om_p$ 
is tuned close to $\om_{\z o}$.  If $\cH^2$ is the intracavity pump power that builds in the $\mu=0$ mode in the quasi-linear regime of operation and at the exact resonance, $\om_{\z o}=\om_p$, then the pump and the finite linewidth effects can be incorporated to the model via  a phenomenological substitution 
\begin{subequations}
	\label{e6:1}
\begin{align}
& i\p_tB_{\mu o}\to  i\p_t B_{\mu o} + i\tfrac{1}{2}\kappa_{\mu o}
\left( B_{\mu o}-\wh\delta_{\mu, \z}\cH e^{i(\omega_{\mu o}-\om_p)t}\right),
\label{e6:1a}\\
& i\p_tB_{\mu e}\to  i\p_t B_{\mu e} + i\tfrac{1}{2}\kappa_{\mu e}
B_{\mu e}
\label{e6:1b}.
\end{align}
\end{subequations}
Here  $\kappa_{\mu s}$ are the linewidth parameters.
$\cH^2$ is expressed via  the pump laser power $\cW$ as
\begin{equation}
\cH^2=\frac{\eta}{\pi}\cF \cW, 
\label{e6:2}
\end{equation}
where $\eta$ is the coupling efficiency into a resonator mode. $\eta=\kappa_{c}/\kappa_{\z 
o}<1$, where $\kappa_{c}$ is the coupling pump rate. $\cF=D_{1o}/\kappa_{\z o}$ is the cavity finesse, or the power 
enhancement factor, which is typically $10^3$ and above for high-quality factor 
microresonators. Theory of power enhancement effects in ring cavities can be found in, e.g., Ref.  \cite{book1}.

Thus, the coupled-mode equations ready to be used for numerical modelling of practical experimental setups are
\begin{subequations}
	\label{e6:3}
	\begin{align}
	 i \p_t B_{\mu o}&=-i\tfrac{1}{2}\kappa_{\mu o}
	\left( B_{\mu o}-\wh\delta_{\mu, \z}\cH e^{i(\omega_{\mu o}-\om_p)t}\right)
	\nn \\
	&-\gamma_{\mu o}
	\sum_{\mu_1\mu_2 m}\wh\delta_{\mu_2,\mu+\mu_1-g-m}
	G_m B^*_{\mu_1 o} B_{\mu_2 e} 
	e^{-i(\om_{\mu_2 e}-\om_{\mu_1 o}-\om_{\mu o})t},
	\label{e6:3a}\\
	i \p_tB_{\mu e}&=-i\tfrac{1}{2}\kappa_{\mu e}	B_{\mu e}
	\nn \\
	&-\gamma_{\mu e}
	\sum_{\mu_1\mu_2 m}\wh\delta_{\mu_2,\mu-\mu_1+g-m}
	G_m B_{\mu_1 o}B_{\mu_2 o}
	e^{	-i(\om_{\mu_2 o}+\om_{\mu_1 o}-\om_{\mu e})t}.
	\label{e6:3b}
	\end{align}
\end{subequations}
Quality factors between $10^9$ and $10^7$
would correspond to   $\kappa_{\mu s}/2\pi$ varying in a 
range from below one to tens of MHz, 
 with the tendency 
towards smaller quality factors for higher frequencies.

\section{Pseudo-spectral form of coupled-mode equations}
Eqs. \bref{e6:3} are now ready to be set as an initial 
value problem and integrated in time with a
suitable solver, including the variable 
step ones. However,  nonlinear sums remain a bottleneck in the multi-mode regime, 
since one should be summing up $\sim N$ exponential 
terms in $2N$ equations. Here we address how this can be handled efficiently using Fourier transform, that gives $N\ln N$ scaling.
Expressing and computing multi-mode Kerr nonlinearity (four-wave mixing)
via Fourier transforms has been discussed in Refs. \cite{skr,optcom}. This is a so-called pseudo-spectral approach. It  originates in solving quantum mechanical  Schrodinger equation with a trapping potential using a basis of the free space eigenstates  \cite{pseudo}.

To achieve pseudo-spectral formulation, we first introduce detunings
$\delta_{\mu o}$ between the laser frequency, $\om_p$, and frequencies of the ordinary 
resonator modes, and $\delta_{\mu e}$ between $2\om_p$ and the extraordinary ones,
\begin{subequations}
	\label{e7:1}
		\begin{align}
&	\delta_{\mu o}=\om_{\mu o}-\om_p,	
\label{e7:1a}\\
& \delta_{\mu e}=\om_{\mu e}-2\om_p.
	\label{e7:1b}
\end{align}
\end{subequations}	
We also define new mode amplitudes, $Q_{\mu s}$,
	\be
 Q_{\mu s}= B_{\mu s}e^{-i\delta_{\mu s}t},
\label{e7:2}
\ee
and the envelope functions $Q_s$,
\be
Q_s=\sum\nolimits_\mu Q_{\mu s}e^{i\mu\ta },~s=o,e.
\label{e7:3}
\ee
New amplitudes and the envelopes are linked as
\be	 
Q_{\mu s}=\int^{2\pi}_0  Q_s e^{-i\mu\ta }\frac{d\ta}{2\pi}.
\label{e7:4}
\ee

Let us note here, that frequency mismatch tensors, 
Eqs. \bref{e5:5}, are straightforwardly re-expressed via detunings:
\begin{subequations}
	\label{e7:5}
	\begin{align}
	\label{e7:5a}	&
	\ep^{eoo}_{\mu_2\mu_1\mu m}=\delta_{\mu_2 e}-\delta_{\mu_1 o}-\delta_{\mu o},\\
	&
	\ep^{ooe}_{\mu_2\mu_1\mu m}=\delta_{\mu_2 o}+\delta_{\mu_1 o}-\delta_{\mu e}.
	\label{e7:5b}
	\end{align}
\end{subequations}
Replacing  $B_{\mu s}$ with $Q_{\mu s}$ in the  nonlinear  
parts of Eqs. \bref{e6:3}  we obtain
\begin{subequations}
	\label{e7:6}
\begin{align}
i \p_t B_{\mu o}&=-i\tfrac{1}{2}\kappa_{\mu o}
\left( B_{\mu o}-\wh\delta_{\mu, \z}\cH e^{i\delta_{\mu o}t}\right)
\nn \\
&
-\gamma_{\mu o}e^{i\delta_{\mu o}t}
\sum_{\mu_1\mu_2 m}\wh\delta_{\mu_2,\mu+\mu_1-g-m}
G_m Q^*_{\mu_1 o} Q_{\mu_2 e},
\label{e7:6a}\\
i \p_t B_{\mu e}&=
-i\tfrac{1}{2}\kappa_{\mu e}B_{\mu e}\nn \\
&-\gamma_{\mu e}e^{i\delta_{\mu e}t}\sum_{\mu_1\mu_2 
m}\wh\delta_{\mu_2,\mu-\mu_1+g-m}
G_m Q_{\mu_1 o}Q_{\mu_2 o}.
\label{e7:6b}
\end{align}
\end{subequations}

Recalling momentum matching conditions, Eqs. \bref{e5:4}, 
we use integral 
representations of the Kronecker $\delta$'s, i.e., 
$\tfrac{1}{2\pi} \int d\ta e^{i(\mu_2-\mu_1+g+m-\mu)\ta}$ and 
$\tfrac{1}{2\pi}\int d\ta e^{i(\mu_2+\mu_1-g+m-\mu)\ta}$, in the sum terms  in 
Eqs. \bref{e7:6}. 
Applying Eqs. \bref{e7:4}, \bref{e4:1} gives
\begin{subequations}
	\label{e7:7}
	\begin{align}
	 i \p_t B_{\mu o}&=-i\tfrac{1}{2}\kappa_{\mu o}
	\left( B_{\mu o}-\wh\delta_{\mu, \z}\cH e^{i\delta_{\mu o}t} \right)\nn \\
	&-\gamma_{\mu o}e^{i\delta_{\mu o}t}
	\int_0^{2\pi} \left(Ge^{ig\ta} Q^*_o Q_e\right) e^{-i\mu\ta}\frac{d\ta}{2\pi},
	\label{e7:7a}\\
		i \p_tB_{\mu e}&=
	-i\tfrac{1}{2}\kappa_{\mu e}B_{\mu e}\nn \\
	&-\gamma_{\mu e}e^{i\delta_{\mu e}t}
	\int_0^{2\pi} \left(Ge^{-ig\ta} Q^2_o\right) e^{-i\mu\ta}\frac{d\ta}{2\pi}.
	\label{e7:7b}
	\end{align}
\end{subequations}

Eqs. \bref{e7:7} 
is the central result of this work. They replace computationally
demanding nonlinear sums of the multiple products of the mode amplitudes with
Fourier transforms of  products of the envelope functions. The 
latter is readily expressed via  Fourier transforms of the amplitudes 
themselves, see Eqs.  \bref{e7:3}. Importantly,  
the real space quasi-phase-matching grating profiles
are incorporated seamlessly and approximation free 
into this approach, and effectively play the role of an axillary field amplitude. 
One could also choose to account only for the leading order phase matching provided by the grating, which is accomplished via a substitution $G(\ta)e^{\pm ig\ta}\to G_{\mp g}$, see Eq. \bref{e4:1}.

\section{Envelope equations}
It is more handy for analytical work, but, at the same time,  
could be computationally more demanding to move to the $Q$-only formulation by  
replacing $B_{\mu s}$ with $Q_{\mu s}$ in the linear terms of Eqs. \bref{e7:7},
\begin{subequations}
	\label{e8:1}
	\begin{align}
	i \p_t Q_{\mu o}&=\delta_{\mu o}Q_{\mu o}-i\tfrac{1}{2}\kappa_{\mu o}
	\left( Q_{\mu o}-\wh\delta_{\mu, \z}\cH  \right)\nn \\ &-\gamma_{\mu o}
	\int_0^{2\pi} \left(Ge^{ig\ta} Q^*_o Q_e\right) e^{-i\mu\ta}\frac{d\ta}{2\pi},
	\label{e8:1a}\\
	i \p_tQ_{\mu e}&=\delta_{\mu e}Q_{\mu e}
	-i\tfrac{1}{2}\kappa_{\mu e}Q_{\mu e}\nn \\ &-\gamma_{\mu e}
	\int_0^{2\pi} \left(Ge^{-ig\ta} Q^2_o\right) e^{-i\mu\ta}\frac{d\ta}{2\pi}.
	\label{e8:1b}
	\end{align}
\end{subequations}
Retaining the grating profile $G(\theta)$ in Eqs. \bref{e8:1} prevents from  
taking the full advantage of working in the 
reference frame rotating with the $D_{1o}$ rate. 
Replacing an integral in  Eqs. \bref{e8:1} with an equivalent $\sum_{\mu_1\mu_2 m}$, one finds that the nonlinear terms are evolving with frequencies given, in the leading order, by  $|\delta_{\mu_1 o}-\delta_{(\mu+\mu_1) e} |\sim D_{1e}\mu$ and $|\delta_{\mu_1 o}+\delta_{(\mu-\mu_1) o})|\sim D_{1o}\mu$. At the same time,  oscillation rates of the nonlinear terms
in Eqs. \bref{e7:7} are  given by $\ep_{\mu_2\mu_1\mu,m}^{s_2s_1s}$, 
see Eqs. \bref{e7:5}. One can readily show that in the leading order those are
$\sim|D_{1e}-D_{1o}|\mu$. For example, $D_{1o}/2\pi\simeq 21$GHz and $D_{1e}/2\pi\simeq 19$GHz in 
a millimetre scale LiNbO$_3$ microresonator  Ref. \cite{jan}. This would give an order of magnitude difference in the time steps needed to achieve the same numerical accuracy if the two systems are solved with the same method and all parameters taken equal. 

To derive the envelope, Lugiato-Lefever 
like, equations, we take  Eqs. \bref{e8:1}, 
neglect by the linewidth dispersion, i.e., $\kappa_{\mu 
s}=\kappa_{\z s}$, and account for dispersion of nonlinearity using 
an approximate Eq. \bref{e5:11}. Transforming back to the physical space  one obtains  a 
system of  the envelope equations with self-steepening terms,
\begin{subequations}
	\label{e8:2}
	\begin{align}
	 i\p_tQ_o=&\delta_{\z o} Q_o+\left(-iD_{1o}\p_\ta 
	-\tfrac{1}{2!}D_{2o}\p_\ta^2+i\tfrac{1}{3!}D_{3o}\p_\ta^3+\dots\right) Q_o
	\nn \\
		-&i\tfrac{1}{2}\kappa_{\z o}\left( Q_o-\cH\right)\nn\\
	-&\gamma_{\z o}\left(1-i\frac{2D_{1e}}{\om_{\z o}}\p_\ta\right) Ge^{ig\ta}Q_o^*Q_e,
	\label{e8:2a}\\
	 i\p_tQ_e=&\delta_{\z e} Q_e+\left(-iD_{1e}\p_\ta 
	-\tfrac{1}{2!}D_{2e}\p_\ta^2+i\tfrac{1}{3!}D_{3e}\p_\ta^3+\dots\right) Q_e
	\nn \\
	-&i\tfrac{1}{2}\kappa_{\z e} Q_e\nn\\
	-&\gamma_{\z e}\left(1-i\frac{D_{1o}}{\om_{\z o}}\p_\ta\right) Ge^{-ig\ta}Q_o^2. 
	\label{e8:2b}
	\end{align}
\end{subequations}
These equations, but without self-steepening, 3rd order dispersion  and 
quasi-phase-matching terms, have been previously used by us in Refs. \cite{jan,soln7,soln4}. Physical values of the self-steepening coefficients are discussed at the end of  Section 5 above.

\section{Nonlinear shift of the pump resonance: chi-2 vs chi-3}
 Here we present a brief evaluation and comparison of the $\chi^{(2)}$ and Kerr induced resonance shifts in microresonators.
An important observation in this regard is that detuning
$\delta_{\z e}$  is controlled on one side by the pump frequency,  
and on the other by phase matching and/or resonator dispersion. Indeed,
\be
\delta_{\z e}=2\delta_{\z o}-\ep,
\label{e9:1}
\ee
where $\ep$ is the cavity resonance frequency mismatch parameter defined in Eq. \bref{e4:3}.
We assume that  phase-matching has been arranged to give $|\ep|$ that is sufficiently large to dominate pump detuning and linewidth and sufficiently small to give a non-negligible second harmonic, i.e., $D_{1s}>|\ep|\gg\kappa_{\z s}$. 
Then, assuming the single mode operation, 
\be
Q_{\z e}\simeq -\frac{\gamma_{\z e}}{\ep}G_gQ_{\z o}^2.
\label{e9:2}
\ee
Substituting Eq. \bref{e9:2} into Eq. \bref{e8:1a}, one finds that  $\chi^{(2)}$ nonlinearity induces an effective, or cascaded \cite{ronald0,rev3}, Kerr effect for the ordinary field, which gives the following nonlinear shift of $\om_{\z o}$ ($G_{\pm g}=1$):  
\be
\Delta_{cas}=\frac{\gamma_{\z o}\gamma_{\z e}}{\ep}|Q_{\z o}|^2\simeq
\frac{\om_{\z o}}{ 2S_{\z o}n_{\z o}}
\frac{16{\cal Z}_{vac}\om_{\z o}d_{31}^2}{9n_{\z o}^2 \ep }|Q_{\z o}|^2.
\label{e9:3}
\ee

Using Ref. \cite{skr}, 
the shift of  $\om_{\z o}$ due to intrinsic Kerr effect is 
\be
\Delta=-\frac{\om_{\z o}}{2S_{\z o}n_{\z o}}n_2 |Q_{\z o}|^2.
\label{e9:4}
\ee
We take the intrinsic Kerr coefficient for LN as
$n_2\simeq 5\cdot 10^{-19}$m$^2$/W, noting that there is some spread 
of values met in the literature.

Comparing Eqs. \bref{e9:3} and \bref{e9:4}, leads to the cascaded, $\chi^{(2)}$ induced, Kerr coefficient
\be
n_2^{cas}=-\frac{16{\cal Z}_{vac}\om_{\z o}d_{31}^2}{9 n_{\z o}^2\ep}.
\label{e9:7}
\ee 
Choosing, as an example, $\ep/2\pi=-1$GHz gives 
$n_2^{cas}\simeq 6\cdot 10^{-16}$m$^2$/W, which is three orders of magnitude above the intrinsic $n_2$. 
For smaller $\ep$, i.e., better phase matching,  dominance of $n_2^{cas}$ will increase according to $n_2^{cas}\sim 1/\ep$.
For $|\ep|\lesssim\kappa_{\z e}$, this scaling should be generalised 
to include saturation of $n_2^{cas}$ with power and detuning effects \cite{soln7}. 
Threshold for the comb generation via intrinsic Kerr nonlinearity in integrated LN resonators is $\lesssim 100$mW and scales inversely with $n_2$ \cite{micro3}. If the cascaded nonlinearity is triggered, then according to $n_2^{cas}/n_2\sim 10^3$, the threshold  is expected to decrease to $100\mu$W and less.
Exploring weighting of $\chi^{(2)}$ vs $\chi^{(3)}$ effects further can easily constitute a separate investigation and we refer a reader to Refs. \cite{alex,loncar,soln7,china}
for more information. Impacts of Raman and slow photorefractive responses 
of LN on comb generation have been analysed in, e.g., \cite{soln6,micro3}.

\section{Summary}
We have  presented an ab-initio derivation of the coupled-mode equations 
describing nonlinear wave mixing processes 
in  microresonators with quadratic nonlinearity and quasi-phase-matching. 
Main features of our coupled-mode formulation given by  Eqs. \bref{e7:7} 
are - (i) nonlinear terms are evaluated pseudo-spectrally, i.e.,
using Fourier transforms of the products of the real space envelope functions; (ii) number of nonlinear coefficients to be calculated to account for dispersion of the nonlinear interaction is optimized, Eqs. \bref{e5:8}; (iii) arbitrary profiles of the quasi-phase-matching gratings naturally enter the pseudo-spectral formulation. 

While we considered in details an example of phase matching between ordinary and extraordinary waves in an LN crystal mediated by the $d_{31}$ coefficient, we also provide a full tensor formulation of the nonlinear response, that  includes all other types of three-wave mixing, see Eqs. \bref{e3:5}, and allows extensions of our approach.
We further demonstrate that the coupled-mode Eqs. \bref{e7:7} with the nonlinear coefficients approximated by Eqs. \bref{e5:11} are formally equivalent to a pair of the envelope equations, Eqs. \bref{e8:2}, with self-steepening terms.
Balance of the time-scales involved suggests that Eqs. \bref{e7:7} are computationally advantageous over Eqs. \bref{e8:2}. Opportunities for future studies 
concerning both physics and numerics related aspects of the 
problem and also the mapping of the models  onto a rich variety of experimental settings are numerous.

\section*{Funding}
EU Horizon 2020 Framework Programme (812818, MICROCOMB); Russian
Science Foundation (17-12-01413).



\end{document}